\documentclass{iau} 

\usepackage{graphicx}
\usepackage{xcolor}
\usepackage[hyphens]{url}
\usepackage{hyperref}
\usepackage[utf8]{inputenc}
\hypersetup{
     colorlinks  = true,
     linkcolor   = blue, 
     anchorcolor = BrickRed,
     citecolor   = blue,
     filecolor   = blue,
     menucolor   = blue,
     runcolor    = cyan,
     urlcolor    = blue
}
\usepackage[authoryear]{natbib}

\pubyear{2019}
\volume{355} 
\jname{The Realm of the Low-Surface-Brightness Universe}
\editors{D. Valls-Gabaud, I. Trujillo \& S. Okamoto, eds.}

\title[Halo Structural Parameters of LSB galaxies] 
{Halo Structural Parameters of Low Surface Brightness Galaxies}

\author[L.E. Pérez-Montaño \& B. Cervantes Sodi]   
{Luis Enrique Pérez-Montaño$^1$
 \and Bernardo Cervantes Sodi$^1$}

\affiliation{$^1$Instituto de Radioastronomía y Astrofísica, Universidad Nacional Autónoma de México. \\ Apdo. Postal 3-72, Morelia, Michoacán, 58089, México. \\ email: {\tt l.perez@irya.unam.mx} \\[\affilskip]}

\begin{document}

\maketitle

\begin{abstract}
We build a volume-limited sample of galaxies derived from the SDSS-DR7 to characterize several physical properties of the dark matter halos where LSB galaxies reside. Using an observational proxy for the assembly time, we found that LSB galaxies assembly half of their total halo mass later than HSB ones, reinforcing the idea of them being unevolved systems. We use 5 different methods to estimate the total halo mass, finding that the total stellar-to-halo mass ratio is up to 22\% lower in LSB galaxies. Finally, in order to estimate the spin parameter, we use a bulge+disk decomposition to obtain the specific angular momentum $j_*$ of the galaxy, a Tully-Fisher relation to estimate the rotation velocity of the disk, and the 5 different estimations of the halo mass  to calculate the spin parameter. We found that the spin of LSB galaxies is 1.2 to 2 times higher than for HSB ones. We compare these results with a control sample that includes kinematic information, taken from the ALFALFA $\alpha$.100 galaxy catalog, allowing us to measure directly the rotation velocity of the disk. The trends in the values of $j_*$ and $\lambda$ are similar to the volume-limited sample.
\keywords{galaxies: haloes - galaxies: fundamental parameters - galaxies: evolution - galaxies: statistics - galaxies: structure}
\end{abstract}

\firstsection 

\section{Introduction}
It is expected for LSB galaxies to be part of a young population of extragalactic objects \citep{Bothun97}, characterized by having low stellar densities due to their large specific angular momentum $j_{*}$, responsible of spreading their stars and gas over larger areas. These galaxies are expected to form in fast rotating halos with high values of the spin parameter $\lambda$ \citep{Jimenez98, Boissier03, KimLee13, Cerv17}. In this work, we provide an observational counterpart to theoretical studies related with the properties of LSB galaxies mentioned above. We first estimate the halo assembly time, defined as the redshift at which the main dark matter (DM) halo has assembled half of its total mass \citep{Wang11}, using an observational proxy found by \citet{Lim16}. We use a large sample of galaxies to compare the total stellar-to-halo mass ratio between LSB and HSB galaxies, considering 5 different and independent estimations using various methodologies. We also use a bulge+disk decomposition to calculate $j_*$, following the models of \citet{RomFall12} for both components. Finally, we provide a prescription to estimate the spin parameter \citep{Peebles71}, using data accessible from observations to show that LSB galaxies have systematically higher values of the spin parameter than HSB ones.

\section{Sample Selection}
We use a number of different catalogues derived from SDSS-DR7 \citep{Abazajian09}. We take as our base catalgue the Korea Institute for Advance Study Value Added Catalogue (KIAS VAC, \citet{Choi10}) to obtain photometric information of the galaxies. We also use the MPA/JHU SDSS database \citep{Kauffmann03,Brinchmann04}, which provides stellar masses and star formation rates, and a bulge+disk decomposition provided by Simard \citep{Simard11}. To identify central and satellite galaxies, we use the Yang et al. \citep{Yang07} group galaxy catalogue, which also provides information on the dark matter halo masses. When available, we include HI kinematic information from the ALFALFA $\alpha$.100 catalogue \citep{Haynes18}, to obtain a direct measurement of the rotation velocity of the disks. 

Here, we consider a galaxy to be LSB if its central surface brightness \citep{Freeman70} is such that $\mu_B > 22.0$ mag arcsec$^{-2}$, measured in the Johnson's $B-$band. To segregate between LSB and HSB galaxies, we follow the relations from \cite{Trachternach06} and \cite{Zhang09} to calculate the central surface brightness in any band $x$ as
\begin{equation}
	\mu_x = m_x + 2.5\log({2\pi\alpha^2 q}) - 10 \log{(1+z)},
	\label{eq:mu}
\end{equation}

where $m$ is the apparent magnitude, $\alpha$ the exponential radius, $q$ the axis ratio and $z$ the redshift. We apply this expression for the $g-$ and $r-$ bands of the SDSS to obtain the surface brightness in the corresponding $B-$ band \citep{Smith02} as

\begin{equation}
	\mu_B = \mu_g + 0.47(\mu_g-\mu_r) + 0.17.
	\label{eq:mu_B}
\end{equation} 

\subsection{Volume-limited sample}
We build this sample following the criteria of previous works \citep{Zhong08, Galaz11}, in which we consider those galaxies brighter than $M_r = -19.8$ mag, within a volume of $0.01<z<0.1$. We only include face-on ($q > 0.4$), late type spiral galaxies ($fracDev<0.9$). Our final sample consists on 64,351 galaxies of which 21,273 are LSBs and 43,078 are HSBs. 

\subsection{Control sample}
This sample includes kinematic information from ALFALFA $\alpha$.100, unfortunately this catalogue does not provide a full map of the sky, so a match with the objects in the volume-limited sample will considerably reduce the number of galaxies. In order to provide a robust statistical study, we build control samples considering galaxies with inclination angles between 25$^\circ$ and 75$^\circ$ (taken from \citet{Simard11}) to guarantee a correct measure of HI data. These control samples consist of pair of galaxies, an LSB with an HSB counterpart, with similar stellar mass ($\Delta \log(M_*) < 0.05$) and redshift ($\Delta z < 0.002$). By doing this, we guarantee that the differences between LSB and HSB galaxies do not depend on $M_*$ and $z$.\\


\section{Methodology}
\label{sec:methods}
For the assembly time $z_f$, we consider the following quantity:
\begin{equation}
	f_c \equiv \frac{M_{*,c}}{M_H},
	\label{eq:assemblytime}
\end{equation}

where $M_{*,c}$ corresponds to the stellar mass of the central galaxy (in our case, considered as the most massive galaxy of a group) and $M_H$ the mass of the main DM halo. It has been shown \citep{Lim16} that $\log(f_c)\propto z_f$ for a wide range of halo mass values ($11.4 \leq \log(M_H) \leq 14.0$). Eq. \ref{eq:assemblytime} will be referred hereafter as \textit{the assembly time} of the halo, and is used as our observational proxy of the real assembly time.

The spin parameter is defined \citep{Peebles71} as a dimensionless quantity that measures the degree of rotational support of a galaxy, and is given by 
\begin{equation}
	\lambda = \frac{L|E|^{1/2}}{G M^{5/2}},
	\label{eq:spin_def}
\end{equation}

where $L$ is the total angular momentum, $E$ the total energy, $G$ the gravitational constant and $M$ the dynamical mass of the galaxy. In our case, the later corresponds to the DM halo mass. In what follows, we implement a simple model to give an estimate of the spin parameter in terms of accessible physical parameters. Assuming that the halo is modelled by an isothermal sphere, and considering the total energy to be dominated by the gravity potential energy, if the halo is virialized, and letting the baryons and DM to be homogeneously mixed, such that any gravitational torque affects both components equally, by conservation of angular momentum, eq. \ref{eq:spin_def} can be re-written as 
\begin{equation}
	\lambda_{obs} = \frac{j_* v_{rot}}{\sqrt{2}GM}.
	\label{eq:spin_obs}
\end{equation}

For the volume-limited sample, we adopt the $v_{rot}$ from a stellar TF relation \citep{Reyes11}, while for control samples, this is obtained directly from ALFALFA. The specific angular momentum of the stellar component is obtained following \citet{RomFall12}, considering a bulge+disk decomposition such that
\begin{equation}
	j_* = f_b j_b + (1-f_b)j_d,
\end{equation}

where $f_b$ is the fraction of the total mass (luminosity\footnote{Assuming a constant mass-luminosity ratio.}) associated to the bulge, taken from \citet{Simard11}.

Finally, we consider 5 different approaches to the DM halo mass (details available in \citet{PerezMontano19}), in order to compute not only the total stellar-to-halo mass fraction, but also the spin parameter by substituting the corresponding value of $M$ in eq. \ref{eq:spin_obs}. These approaches are:
\begin{itemize}
	\item Constant stellar-to-halo mass ratio \citep{HdzCerv06}.
	\item Surface density-scaled ratio \citep{Gnedin07}.
	\item $v_{rot}-v_H$ relation obtained from \citet{Pap11}.
	\item Halo occupation model \citep{Hudson15}.
	\item Galaxy group catalog from \citet{Yang07}.
\end{itemize}
 
 The different values of $\lambda$ are compared with an alternative expression \citep{Meurer18}, in which the spin parameter does not depend on the value of $M_H$, but is a function of the orbital time $t_{orb}$:
\begin{equation}
	\lambda = \frac{\sqrt{50}}{\pi} \frac{t_{orb}(R)}{t_H} \frac{R_d}{R}
	\label{eq:spin_clocks}
\end{equation} 

\section{Results}
\begin{figure}
	\begin{tabular}{cc}
	\includegraphics[width=0.54\textwidth]{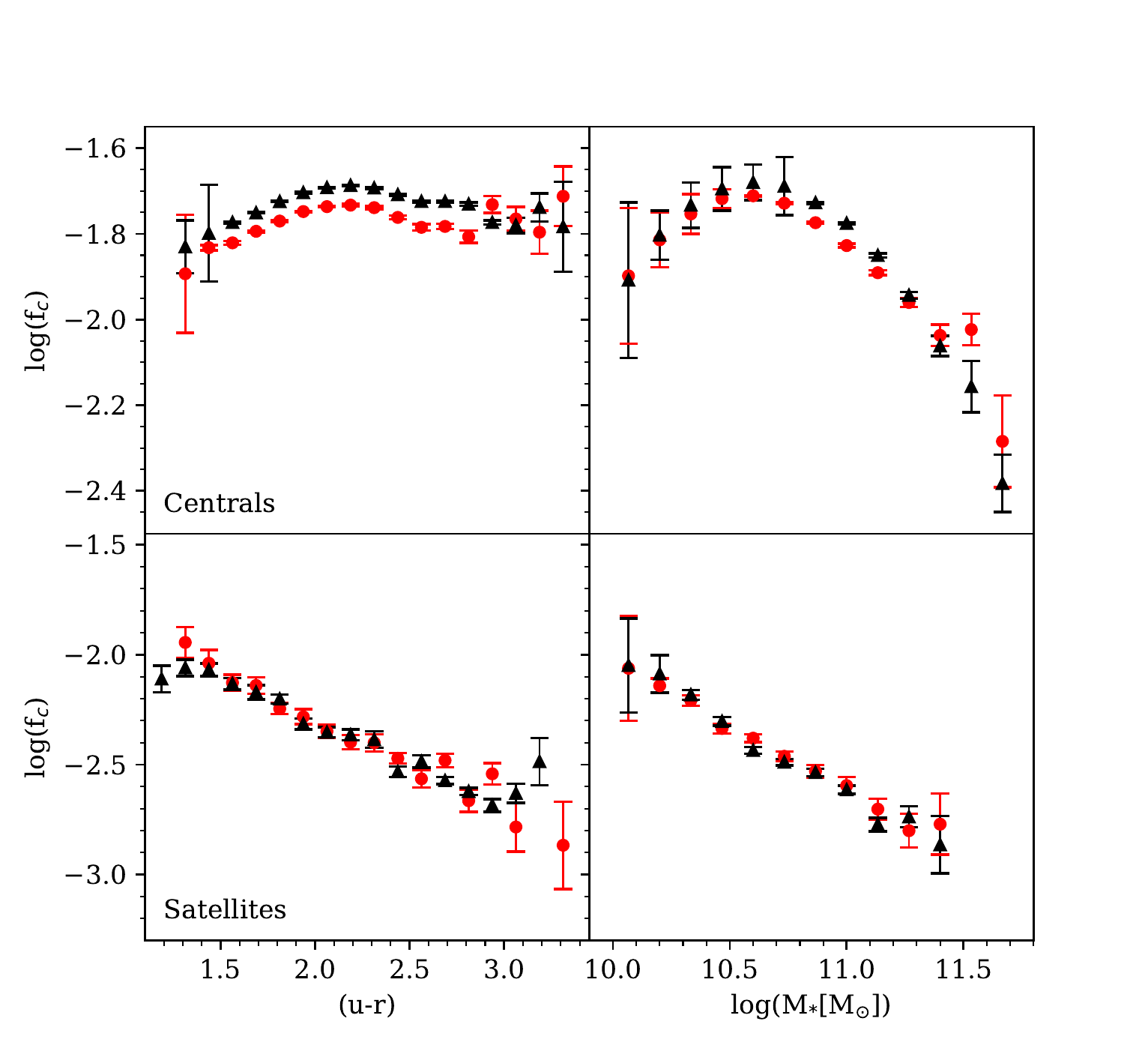} & \includegraphics[width=0.5\textwidth]{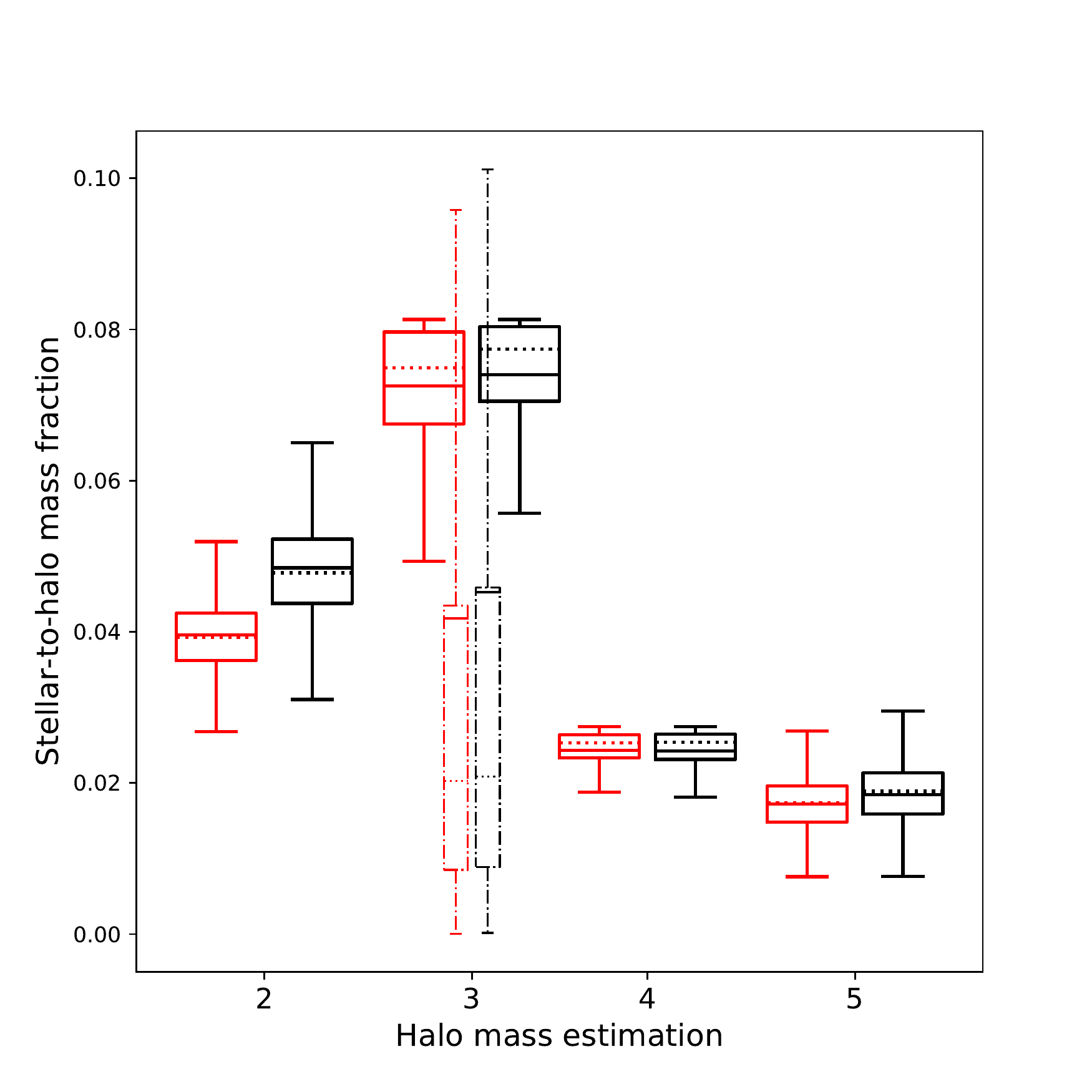} \\
	\label{fig:fc_AND_MsMh}
	\end{tabular}
	\caption{\textit{Left}: Assembly time as a function of colour and stellar mass, for central and satellite galaxies. Larger values of $f_c$ correspond to earlier assembly times. We note that, particularly for centrals, not only the stars, but also de DM halo is a younger structure compared with HSB ones. \textit{Right}: Box plots enclosing 50\% of the data, whiskers represent 3/5 of the IQR. Independently of the estimation employed, LSB galaxies are more DM dominated than HSB ones. In both plots, red (black) colored symbols correspond to LSBs (HSBs). This colour convention will be adopted through this paper.}
\end{figure}

The left panel of fig. \ref{fig:fc_AND_MsMh} shows that, for central galaxies, the halos of LSB galaxies are assembled at latter epochs. This is an important result, since it is well known that the stellar populations of LSB galaxies are younger than the HSB ones. In this regard, this is the first time that it is shown that the halo of LSB galaxies is also formed later when compared with HSBs. With this general picture, we can consider LSB galaxies as young, unevolved objects. In the case of satellite galaxies, we do not find a clear difference.

For the stellar-to-halo mass fraction, we apply the five methods mentioned above to compute $f_n=M_*/M_H$, where the sub index $n$ indicates the corresponding DM halo mass estimation. In the right panel of fig.\ref{fig:fc_AND_MsMh}, we make box plots enclosing 50\% of the data. The solid line-type boxes indicate $M_*/M_H$ for the volume-limited sample, whereas dash-dotted line boxes correspond to the $f_3$ for the control sample\footnote{$f_3$ is the only DM estimation that depends on the value of $v_{rot}$} (see references in sec. \ref{sec:methods} for details). Inside each plot, solid and dashed lines correspond to the mean and median values of each distribution respectively. Kolmogorov-Smilnov (KS) tests are applied to compare between LSBs and HSBs distributions, obtaining in all cases $p$-values lower than 0.001, which allows us to reject our null-hypothesis that both sub-samples came from the same parent distribution. Therefore, this can be interpreted as two different populations of objects. It is important to note that, independently of the estimation, we found that LSB galaxies have a lower total stellar-to-halo mass ratio than HSBs (up to 22\%).

\begin{figure}
\centering
	\begin{tabular}{cc}
	\includegraphics[width=0.4\textwidth]{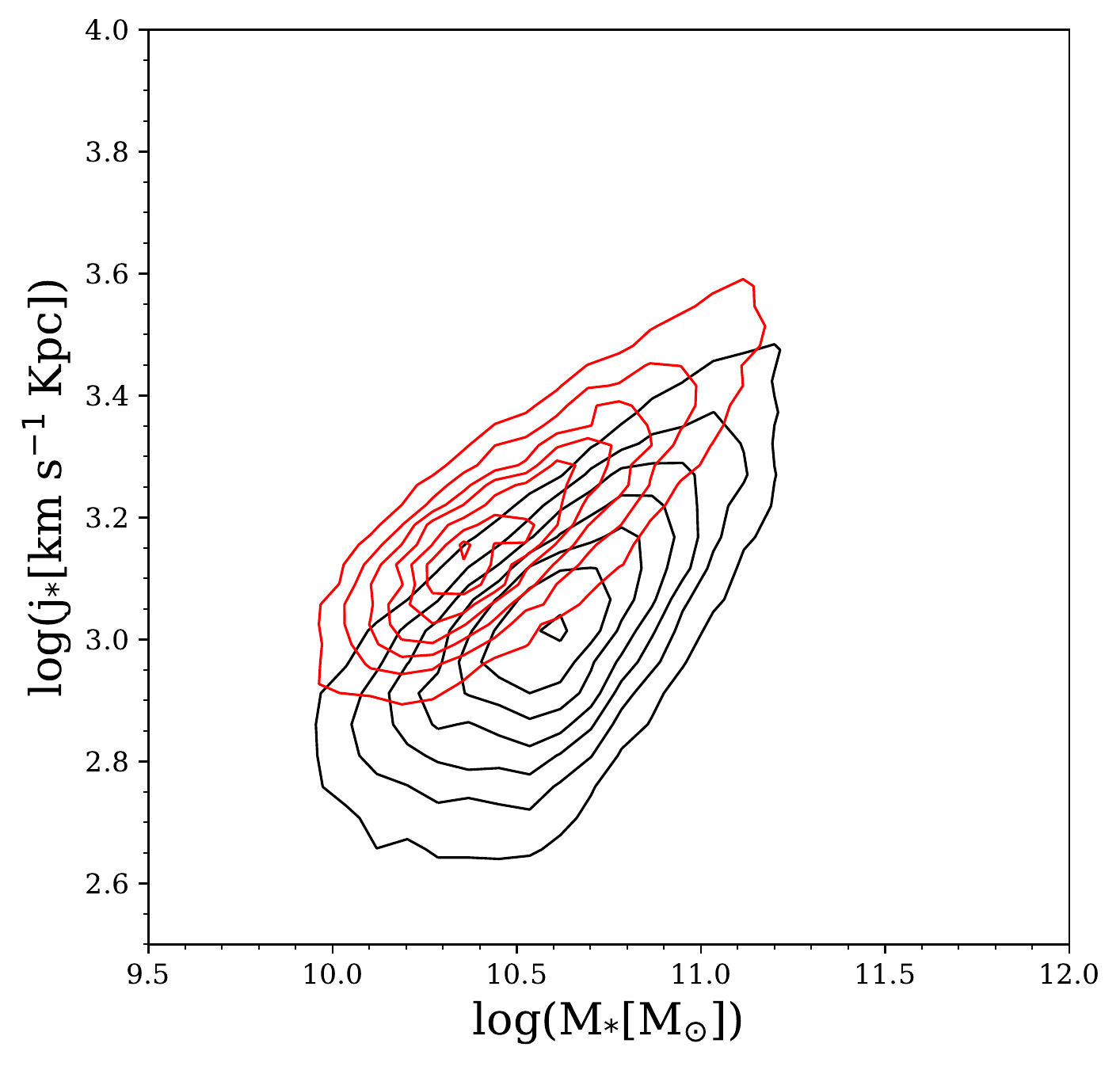} & \includegraphics[width=0.4\textwidth]{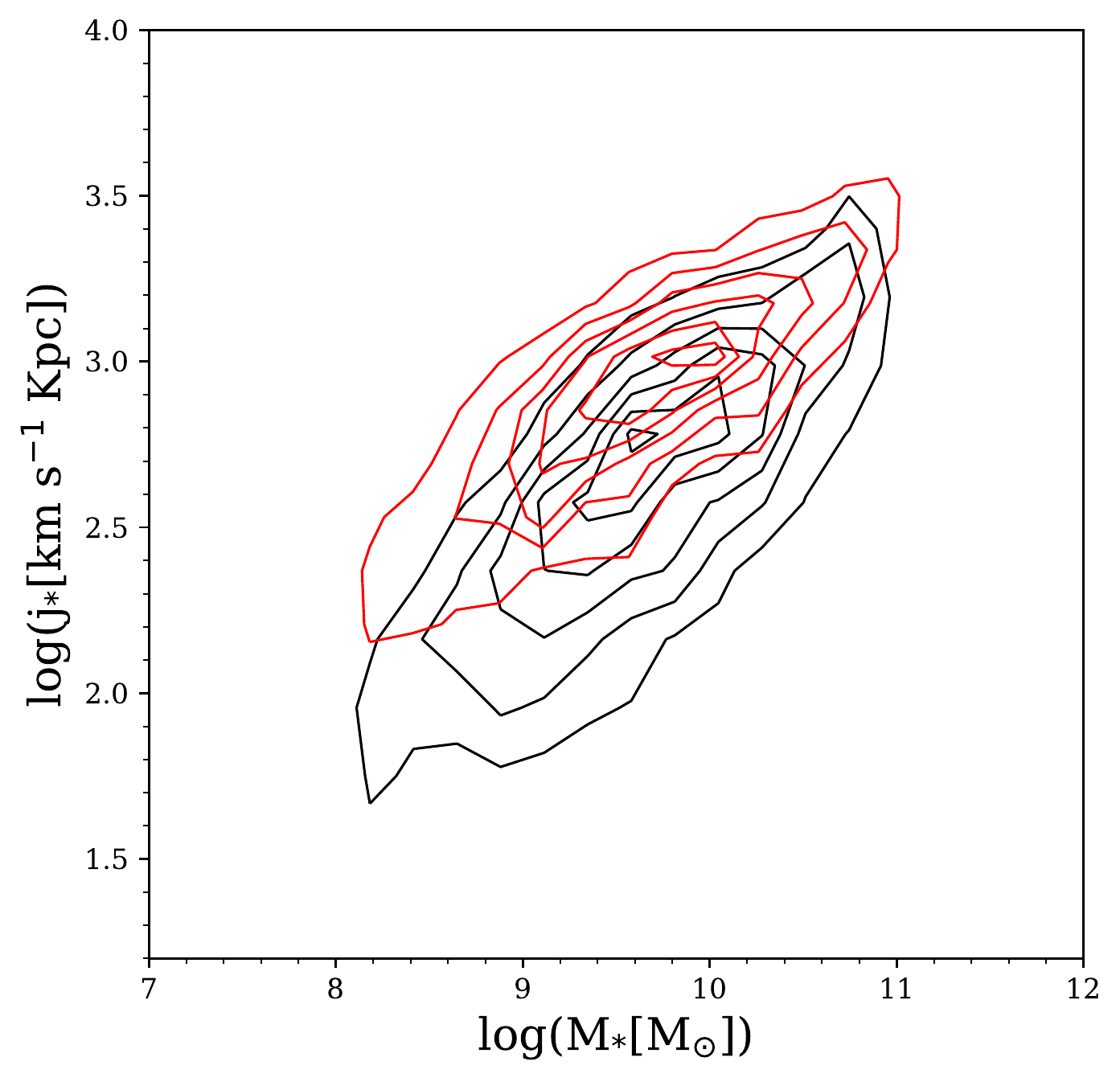} \\
	\includegraphics[width=0.4\textwidth]{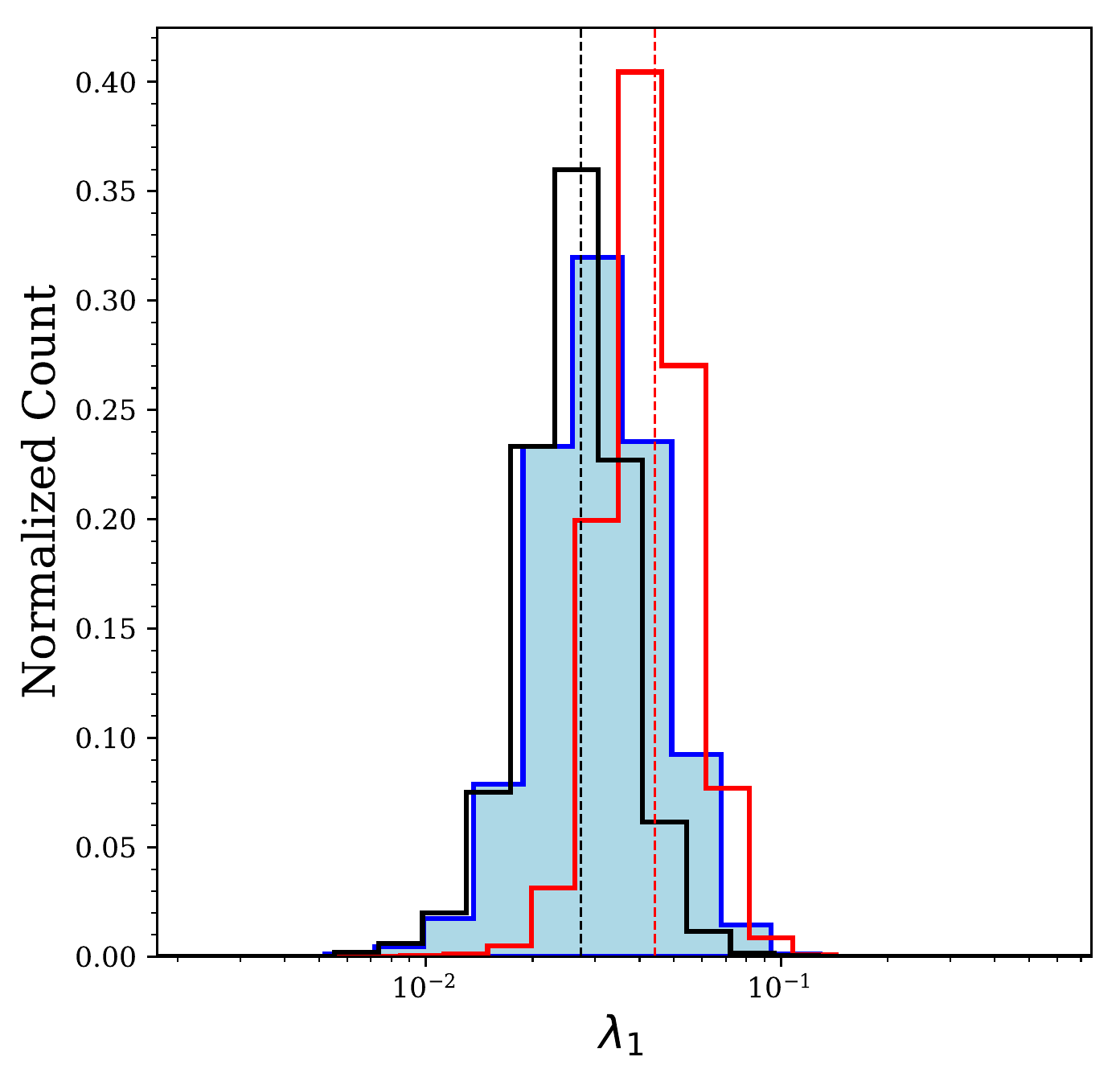} & \includegraphics[width=0.4\textwidth]{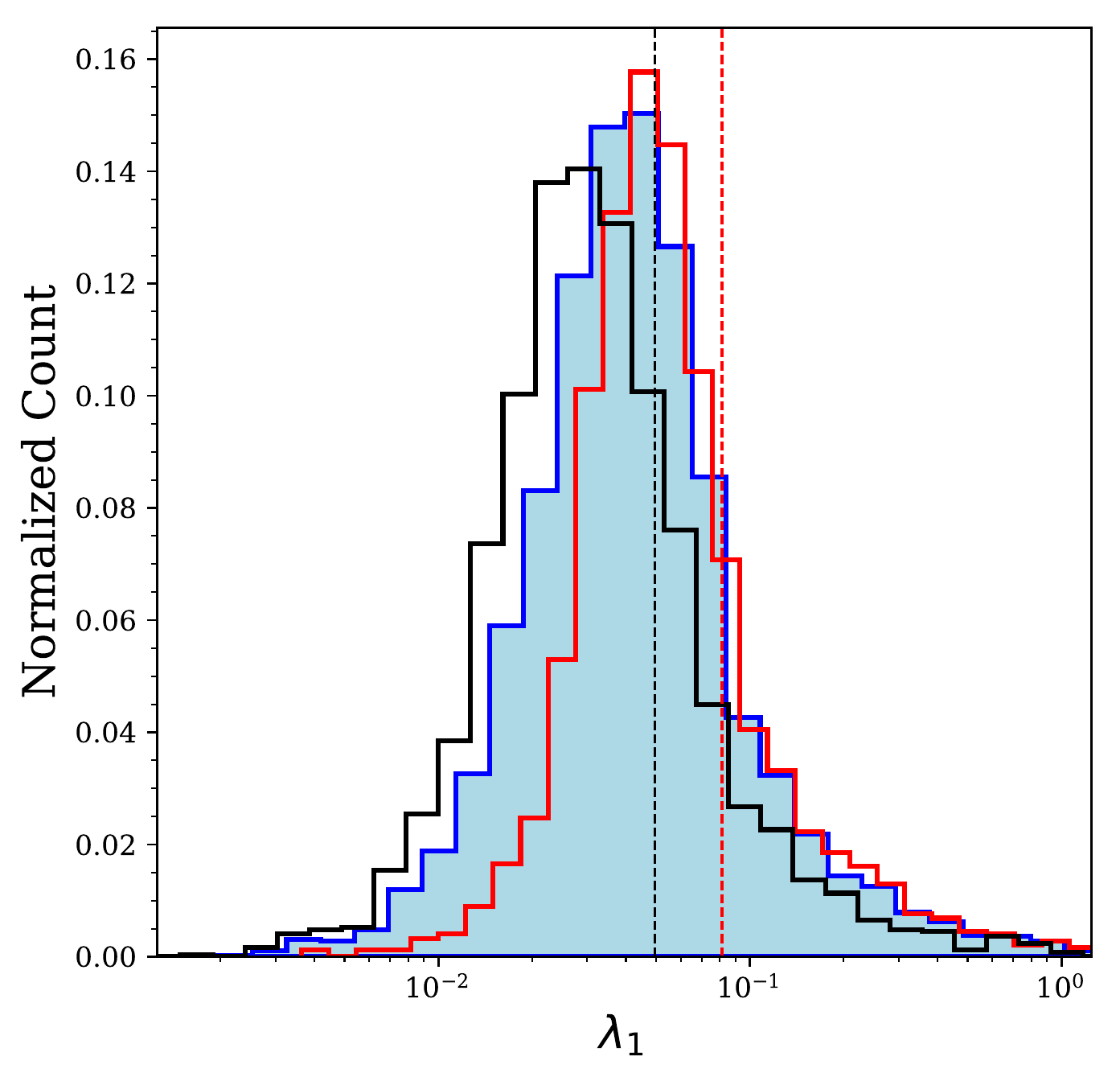}\\
	\label{fig:j_AND_spin}
	\end{tabular}
	\caption{\textit{Top}: Specific angular momentum as a function of stellar mass for volume limited (left) and control samples (right). In both cases, for a fixed stellar mass, LSB galaxies show higher values of $j_*$ when compared with HSBs. \textit{Bottom}: Spin parameter distribution for Volume Limited (left) and control samples (right). Dashed lines represent the mean value of the distribution, whereas blue-coloured histogram represent the distribution of $\lambda$ for the full volume-limited or control samples. In both cases, LSB galaxies present systematically higher values of $\lambda$. This behavior is similar for all the estimations of $\lambda$ \citep{PerezMontano19}. }
\end{figure}

We then compute the specific angular momentum of the stellar component following the models of Romanowsky \& Fall (2012). Top panels of figure \ref{fig:j_AND_spin} show contour plots of the data in the volume-limited (left) and control samples (right). Each contour encloses $\sim$15\% of the data. We note that LSB galaxies present systematically higher values, this implies that the stars would spread over a larger area, and this may be the reason of their low surface brightness. Besides, the gas would not reach critical densities to form stars. 

Finally, we use the 5 different approaches to the DM halo mass from eq. \ref{eq:spin_obs} to compute the spin parameter, and we compare them with the spin obtained with eq. \ref{eq:spin_clocks}. In all the cases, and independently of the method, the different distributions for $\lambda$ show the same tendencies: for LSB galaxies, the spin is larger than for HSBs galaxies \citep{PerezMontano19}, confirming our initial hypothesis and previous theoretical results. For volume limited samples, we obtain that $1.3 < \lambda_{LSB} / \lambda_{HSB} < 1.7$, whereas for control samples, $1.65 < \lambda_{LSB} / \lambda_{HSB} < 2.0$\footnote{In both cases, this number corresponds to the ratio between the main value of  $\lambda$ for LSBs and HSBs.}. As with the stellar-to-halo mass fraction, these tendencies are the same independently of the method employed. KS-tests give again $p$-values $<$ 0.001.

\section{Summary and Conclusions}
Using a large sample of galaxies derived from SDSS-DR7, we build a volume-limited sample comparable with previous studies of LSB galaxies, and we complement this with control samples in order to include kinematic information of ALFALFA $\alpha$.100, to study differences of the halo assembly times, the stellar-to-halo mass ratios, the specific angular momentum and the spin parameter between LSB and HSB galaxies.

We find that LSB galaxies assemble half of their total halo mass at latter epochs when compared with HSBs, supporting the common knowledge that these systems are young objects in formation process. Using 5 different estimators for $f=M_*/M_H$, we observe that LSB galaxies have lower values of $f$, with differences reaching up to 22\% between them. We also found that the specific angular momentum $j_*$ is larger for LSBs than HSBs, and this is the main reason of their lower stellar (and gas) densities. Finally, using our 5 different values of $M_H$ we compute the spin parameter together with an alternative expression that does not depend directly on the mass of the DM halo, and found that $\lambda$ is $\sim$ 1.3-2.0 larger in LSBs when compared with HSBs. These differences could arise also if HSB galaxies retain less angular momentum, reflected on a higher bulge-to-disk ratio \citep{FallRom13}.

\begin{discussion}
\end{discussion}


\begin{thebibliography}{99}

\bibitem[Abazajian et al. (2009)]{Abazajian09} Abazajian, K.~N., Adelman-McCarthy, J.~K., Ag{\"u}eros, M.~A., et al.\ 2009, ApJS, 182, 543

\bibitem[Boissier et al. (2003)]{Boissier03} Boissier, S., Monnier Ragaigne, D., van Driel, W., Balkowski, C., \& Prantzos, N.\ 2003, ApSS, 284, 913

\bibitem[Bothun, Impey \& McGaugh (1997)]{Bothun97} Bothun G., Impey C., McGaugh S., 1997, PASP, 109, 745

\bibitem[Brinchmann et al. (2004)]{Brinchmann04}Brinchmann, J., Charlot, S., White, S.~D.~M., et al.\ 2004, MNRAS, 351, 1151

\bibitem[Choi et al. (2010)]{Choi10} Choi, Y.-Y., Han, D.-H., \& Kim, S.~S.\ 2010, Journal of Korean Astronomical Society, 43, 191

\bibitem[Cervantes Sodi \& S{\'a}nchez Garc{\'{\i}}a(2017)]{Cerv17} Cervantes Sodi, B., \& S{\'a}nchez Garc{\'{\i}}a, O.\ 2017, ApJ, 847, 37 

\bibitem[de Block, McGaugh \& Rubin (2001)]{DeBlockMcGaughRubin01} de Block, W.,J.,G., McGaugh S., Rubin, V., \ 2001, AJ, 122, 2396

\bibitem[Fall \& Romanowsky (2013)]{FallRom13} Fall, S.~M., \& Romanowsky, A.~J.\ 2013, ApJL, 769, L26

\bibitem[Freeman (1970)]{Freeman70} Freeman, K.C. \ 1970, ApJ, 160, 811

\bibitem[Galaz et al. (2011)]{Galaz11} Galaz, G., Herrera-Camus, R., Garcia-Lambas, D., \& Padilla, N.\ 2011, ApJ, 728, 74

\bibitem[Genel et al. (2015)]{Genel15} Genel, S., Fall, S.~M., Hernquist, L., et al.\ 2015, ApJL, 804, L40

\bibitem[Gnedin et al. (2007)]{Gnedin07} Gnedin, O.~Y., Weinberg, D.~H., Pizagno, J., Prada, F., \& Rix, H.-W.\ 2007, ApJ, 671, 1115

\bibitem[Haynes et al.(2018)]{Haynes18} Haynes, M.~P., Giovanelli, R., Kent, B.~R., et al.\ 2018, ApJ, 861, 49 

\bibitem[Hernández \& Cervantes-Sodi (2006)]{HdzCerv06} Hernandez, X., \& Cervantes-Sodi, B.\ 2006, MNRAS, 368, 351 

\bibitem[Hudson et al.(2015)]{Hudson15} Hudson, M.~J., Gillis, B.~R., Coupon, J., et al.\ 2015, MNRAS, 447, 298

\bibitem[Jimenez et al. (1998)]{Jimenez98} Jimenez, R., Padoan, P., Matteucci, F., \& Heavens, A.~F.\ 1998, MNRAS, 299, 123 

\bibitem[Kauffmann et al. (2003)]{Kauffmann03} Kauffmann, G., Heckman, T.~M., White, S.~D.~M., et al.\ 2003, MNRAS, 341, 33

\bibitem[Kim \& Lee (2013)]{KimLee13} Kim, J. \& Lee, J. \ 2013, MNRAS, 432, 1701

\bibitem[Klypin et al. (2011)]{Klypin11} Klypin, A.~A., Trujillo-Gomez, S., \& Primack, J.\ 2011, ApJ, 740, 102 

\bibitem[Lim et al. (2016)]{Lim16} Lim, S.~H., Mo, H.~J., Wang, H., \& Yang, X.\ 2016, MNRAS, 455, 499

\bibitem[Meurer et al. (2018)]{Meurer18} Meurer, G.~R., Obreschkow, D., Wong, O.~I., et al.\ 2018, MNRAS, 476, 1624 

\bibitem[Papastergis et al. (2011)]{Pap11} Papastergis, E., Martin, A.~M., Giovanelli, R., \& Haynes, M.~P.\ 2011, ApJ, 739, 38 

\bibitem[Peebles (1971)]{Peebles71} Peebles, P.~J.~E.\ 1971, A\&A, 11, 377 

\bibitem[P{\'e}rez-Monta{\~n}o \& Cervantes Sodi (2019)]{PerezMontano19} P{\'e}rez-Monta{\~n}o L.~E. \& Cervantes Sodi B., 2019, MNRAS, accepted (arXiv:1910.03078)

\bibitem[Reyes et al. (2011)]{Reyes11} Reyes, R., Mandelbaum, R., Gunn, J.~E., Pizagno, J., \& Lackner, C.~N.\ 2011, MNRAS, 417, 2347 

\bibitem[Romanowsky \& Fall (2012)]{RomFall12} Romanowsky, A.~J., \& Fall, S.~M.\ 2012, ApJS, 203, 17

\bibitem[Rosebaum et al. (2009)]{Rosenbaum09} Rosenbaum, S.~D., Krusch, E., Bomans, D.~J., \& Dettmar, R.-J.\ 2009, A\&A, 504, 807 

\bibitem[Simard et al.(2011)]{Simard11} Simard, L., Mendel, J.~T., Patton, D.~R., Ellison, S.~L., \& McConnachie, A.~W.\ 2011, ApJS, 196, 11 

\bibitem[Smith et al. (2002)]{Smith02} Smith, J.~A., Tucker, D.~L., Kent, S., et al.\ 2002, AJ, 123, 2121

\bibitem[Trachternach et al.(2006)]{Trachternach06} Trachternach, C., Bomans, D.J., Haberzetti, L. \& Dettmar, R.-J. \ 2006 A\&A, 458, 341

\bibitem[Wang et al. (2011)]{Wang11} Wang, H., Mo, H.~J., Jing, Y.~P., Yang, X., \& Wang, Y.\ 2011, MNRAS, 413, 1973

\bibitem[Yang et al. (2007)]{Yang07} Yang, X., Mo, H.~J., van den Bosch, F.~C., et al.\ 2007, ApJ, 671, 153 

\bibitem[Zhang et al. (2009)]{Zhang09} Zhang, Y., Yang, X., Faltenbacher, A., et al.\ 2009, ApJ, 706, 747

\bibitem[Zhong et al. (2008)]{Zhong08} Zhong, G.~H., Liang, Y.~C., Liu, F.~S., et al.\ 2008, MNRAS, 391, 986 

\end{thebibliography}
\end{document}